\PassOptionsToPackage{table,xcdraw,usenames,dvipsnames}{xcolor}
\documentclass{article}

\usepackage{microtype}
\usepackage{graphicx}
\usepackage{subfigure}
\usepackage{booktabs} % for professional tables

\usepackage{hyperref}

\usepackage[accepted]{icml2025}

\usepackage{amsmath}
\usepackage{amssymb}
\usepackage{mathtools}
\usepackage{amsthm}

\usepackage[capitalize,noabbrev]{cleveref}
\usepackage[ruled,algo2e]{algorithm2e}
\renewcommand{\algorithmicrequire}{\textbf{Input:}}  % Use Input in the format of Algorithm
\renewcommand{\algorithmicensure}{\textbf{Output:}} % Use Output in the format of Algorithm
\SetKwInOut{Input}{Input}\SetKwInOut{Output}{Output}
\theoremstyle{plain}
\newtheorem{theorem}{Theorem}[section]

\theoremstyle{definition}
\newtheorem{definition}[theorem]{Definition}

\theoremstyle{remark}

\usepackage[textsize=tiny]{todonotes}

\usepackage{xspace}
\newcommand{\ourmethod}{{\fontfamily{lmtt}\selectfont \textbf{G-Designer}}\xspace}
\newcommand{\llmname}[1]{{\fontfamily{pcr}\selectfont {#1}}\xspace}

\usepackage[normalem]{ulem}
\usepackage{bbm}
\usepackage{bbding}

\definecolor{ForestGreen}{RGB}{34,139,34}
\definecolor{myyellow}{RGB}{181, 181, 27}

\newcommand{\blue}[1]{$_{\color{BlueGreen}\downarrow #1}$}
\newcommand{\red}[1]{$_{\color{RedOrange}\uparrow #1}$}
\definecolor{darksalmon}{rgb}{0.91, 0.59, 0.48}
\definecolor{emerald}{rgb}{0.31, 0.78, 0.47}
\definecolor{green(pigment)}{rgb}{0.0, 0.65, 0.31}
\definecolor{amaranth}{rgb}{0.9, 0.17, 0.31}
\definecolor{iris}{rgb}{0.35, 0.31, 0.81}
\definecolor{uu}{rgb}{0.95, 0.51, 0.51}
\definecolor{spirodiscoball}{rgb}{0.06, 0.75, 0.99}

\usepackage{multirow}
\usepackage{makecell}
\usepackage{enumerate}
\usepackage{enumitem}
\usepackage{pifont}

\icmltitlerunning{G-Designer}

\begin{document}

\twocolumn[
\icmltitle{\ourmethod: Architecting Multi-agent Communication\\ Topologies via Graph Neural Networks}

% It is OKAY to include author information, even for blind
% submissions: the style file will automatically remove it for you
% unless you've provided the [accepted] option to the icml2025
% package.

% List of affiliations: The first argument should be a (short)
% identifier you will use later to specify author affiliations
% Academic affiliations should list Department, University, City, Region, Country
% Industry affiliations should list Company, City, Region, Country

% You can specify symbols, otherwise they are numbered in order.
% Ideally, you should not use this facility. Affiliations will be numbered
% in order of appearance and this is the preferred way.
\icmlsetsymbol{equal}{*}

\begin{icmlauthorlist}
\icmlauthor{Guibin Zhang}{equal,cuhk,tongji}
\icmlauthor{Yanwei Yue}{equal,tongji}
\icmlauthor{Xiangguo Sun}{cuhk}
\icmlauthor{Guancheng Wan}{emory}
\icmlauthor{Miao Yu}{ustc}\\
\icmlauthor{Junfeng Fang}{nus}
\icmlauthor{Kun Wang}{ustc}
\icmlauthor{Tianlong Chen}{unc}
\icmlauthor{Dawei Cheng}{tongji}
%\icmlauthor{}{sch}
%\icmlauthor{}{sch}
\end{icmlauthorlist}

\icmlaffiliation{cuhk}{CUHK}
\icmlaffiliation{tongji}{Tongji University}
\icmlaffiliation{emory}{Emory University}
\icmlaffiliation{ustc}{USTC}
\icmlaffiliation{unc}{UNC-Chapel Hill}
\icmlaffiliation{nus}{NUS}

\icmlcorrespondingauthor{Guibin Zhang}{guibinz@outlook.com}

% You may provide any keywords that you
% find helpful for describing your paper; these are used to populate
% the "keywords" metadata in the PDF but will not be shown in the document
\icmlkeywords{Machine Learning, ICML}

\vskip 0.3in
]

% this must go after the closing bracket ] following \twocolumn[ ...

% This command actually creates the footnote in the first column
% listing the affiliations and the copyright notice.
% The command takes one argument, which is text to display at the start of the footnote.
% The \icmlEqualContribution command is standard text for equal contribution.
% Remove it (just {}) if you do not need this facility.

%\printAffiliationsAndNotice{}  % leave blank if no need to mention equal contribution
\printAffiliationsAndNotice{\icmlEqualContribution} % otherwise use the standard text.

\begin{abstract}
% \vspace{-5em}
Recent advancements in large language model (LLM)-based agents have demonstrated that collective intelligence can significantly surpass the capabilities of individual agents, primarily due to well-crafted inter-agent communication topologies. Despite the diverse and high-performing designs available, practitioners often face confusion when selecting the most effective pipeline for their specific task: \textit{Which topology is the best choice for my task, avoiding unnecessary communication token overhead while ensuring high-quality solution?} In response to this dilemma, we introduce \ourmethod, an adaptive, efficient, and robust solution for multi-agent deployment, which dynamically designs task-aware, customized communication topologies. Specifically, \ourmethod models the multi-agent system as a multi-agent network, leveraging a variational graph auto-encoder to encode both the nodes (agents) and a task-specific virtual node, and decodes a task-adaptive and high-performing communication topology. Extensive experiments on six benchmarks showcase that \ourmethod is: \textbf{(1) high-performing}, achieving superior results on MMLU with accuracy at $84.50\%$ and on HumanEval with pass@1 at $89.90\%$; \textbf{(2) task-adaptive}, architecting communication protocols tailored to task difficulty, reducing token consumption by up to $95.33\%$ on HumanEval; and \textbf{(3) adversarially robust}, defending against agent adversarial attacks with merely $0.3\%$ accuracy drop. The code is available at \url{https://github.com/yanweiyue/GDesigner}.
\end{abstract}

\vspace{-0.7em}
\section{Introduction}
\vspace{-0.3em}
% Web data, as a naturally occurring data structure, prevails in social networks~\cite{sun2023self,greene2010tracking}, trade networks~\cite{serrano2003topology,fagiolo2010evolution,garlaschelli2007interplay}, transportation systems~\cite{bell1997transportation,farahani2013review}, and recommendation platforms~\citep{fan2019graph,wang2019kgat}, \textit{etc}. Web data can inherently be represented as graphs, where nodes and edges capture the topological relationships between numerous instances. Recently, there has been a surge of interest in the academic community toward optimizing the topology design for Large Language Model-based multi-agent (LLM-MA) systems, essentially, how to \textit{weave the web of agents}~\citep{chen2024internet}.

An LLM-based agent, which integrates the language generation capabilities of LLMs with decision-making and action-execution functionalities \citep{autogpt,babyagi,agentgpt}, has exhibited impressive performance across a wide range of tasks, from reasoning \citep{yao2023react} and code generation \citep{reflexion} to even more complex applications like video gaming \citep{voyager} and autonomous driving \citep{jin2023surrealdriver}. Even more exciting, researchers have discovered that combining multiple LLM-based agents--whether implicitly or explicitly--into a team can outperform individual agents when tackling complex tasks \citep{arXiv2023_MultiAgent-Debate, arXiv2023_MultiAgent-Debate_2, multi-persona, blender, reflexion, PHPrompting, autogen}, demonstrating a form of collaborative intelligence reminiscent of human teamwork in multi-agent systems~\citep{zhang2023exploring}. This emergence of human-esque collective intelligence is fundamentally driven by the design of their topology, \textit{i.e.}, how multi-agents are \textit{connected}, and how they \textit{transmit}, \textit{exchange}, and \textit{assimilate} information reciprocally.

\begin{figure}[!t]
  \centering
  \includegraphics[width=1\linewidth]{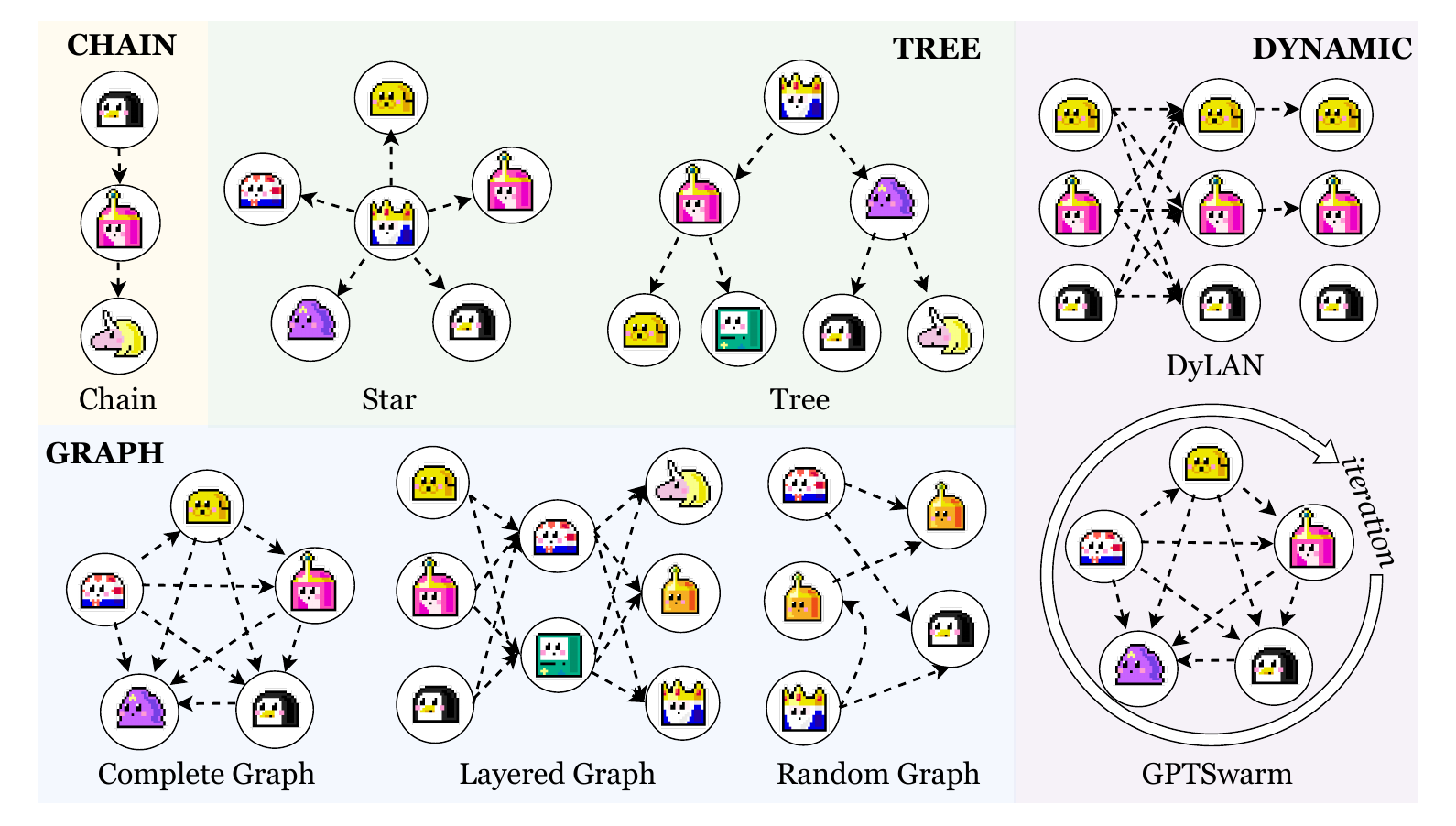}
  \vspace{-2em}
  \caption{Existing practices for LLM-based multi-agent communication topology design. }
   \label{fig:intro_topo}
   \vspace{-1.3em}
\end{figure}

\begin{figure}[!t]
  \centering
  \includegraphics[width=1\linewidth]{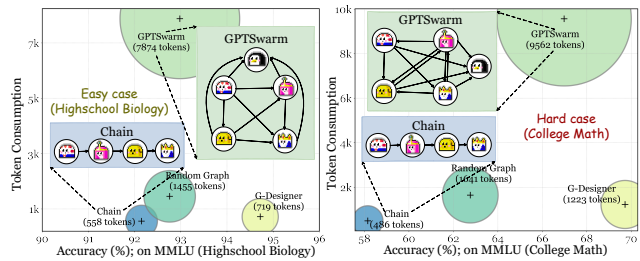}
  \vspace{-1.9em}
  \caption{The token consumption and accuracy of different multi-agent protocols on two subsets of MMLU dataset, ``Highschool Biology'' and ``College Math'', tested with four \texttt{gpt-4}-based agents.}
   \label{fig:intro_scatter}
   \vspace{-1.9em}
\end{figure}

In practice, prior research has extensively explored how multiple instances of LLMs, referred to as agents~\citep{FCS2024_Survey-Agent,arXiv2023_Survey-Agent_2,arXiv2023_Survey-Agent_3,arXiv2024_Survey-Agent_4}, should be structured and organized to converse, collaborate, debate, or even compete. Various topological designs have been investigated, such as chain ~\citep{cot,meta-gpt}, tree~\citep{tot,autogen}, star~\citep{autogen}, complete graphs~\citep{qian2024scaling}, random graphs~\citep{qian2024scaling}, optimizable graphs~\citep{zhuge2024gptswarm,zhang2024cut}, and LLM-based networks~\citep{chatllm-network,arXiv2023_Dynamic-LLM-Agent}. These elaborately designed communication topologies have demonstrated remarkable performance with minimal human supervision, bridging the gap between individual and collective intelligence. Faced with numerous structures available, an inquisitive practitioner might ask: \textit{how should I select or design a topology that best suits my task at hand?}

The question posed above is \textit{non-trivial} and, at times, \textit{perplexing}. A piece of experimental evidence is presented in \Cref{fig:intro_scatter}, where we evaluated the performance of different multi-agent structures on the MMLU dataset~\citep{mmlu}, a collection of multiple-choice questions across various subjects. The results reveal that even within the same dataset, the suitability of different communication topologies varies. \ding{182} \textbf{Simpler Case}: in the simpler "High School Biology" subset, the chain structure performs comparably to the complex GPTSwarm, while consuming significantly fewer tokens (0.5k versus 7.8k). In this case, the chain structure is clearly a more economical choice. \ding{183} \textbf{Harder Case}: However, for the more challenging "College Mathematics" subset, GPTSwarm outperforms the chain structure by $8.75\%$, primarily attributed to its intricate topology and prompt optimization. In summary, practitioners often find it challenging to \textit{effortlessly identify the most efficient and complexity-adaptive multi-agent topology for a given task}.

In light of this dilemma, we propose the \textbf{LLM-based \uwave{M}ulti-\uwave{a}gent \uwave{C}ommunication \uwave{P}rotocol (MACP)}, establishing standardized guidance for LLM-MA topology design:

\noindent\fbox{%
    \parbox{0.97\linewidth}{%
       \textbf{Multi-agent Communication Protocol (MACP)}: \textit{Given a task/query $q$, an optimal LLM-MA communication topology for $q$ should satisfy the following protocol logics: \textbf{(1) Effectiveness}: The communication structure must effectively produce the qualified solution for $q$; \textbf{(2) Complexity-adaptiveness}: The topology should dynamically adjust to the complexity of the task, minimizing communication overhead; \textbf{(3) Adversarial robustness}: The topology should maintain reliable under adversarial attacks.}
    }\label{box:macp}
}

The formal definition of MACP is provided in \Cref{sec:protocol}. To design a communication topology that ideally adheres to the MACP principles, we propose \textit{an effective, adaptive, and robust LLM-powered multi-agent communication graph designer}, termed \ourmethod. Technically, \ourmethod first architects a multi-agent graph, where each agent, along with its specific properties (e.g., profile~\citep{NeurIPS2023_Agent-SoM}, external API tools~\citep{zhuang2023toolchain}, or knowledge base~\citep{chen2024benchmarking}), is represented as a node, and communication between agents forms the edges. \ourmethod employs a variational graph auto-encoder to encode the nodes (agents) along with task-specific information, and to decode the resulting collaboration network between agents. This input-dependent paradigm allows \ourmethod to 
%gauge task complexity and 
design \textbf{task-adaptive, high-performing communication topology}, which is, at the same time, assured of efficiency and robustness with sparsity regularization. Unlike previous LLM-based multi-agent topology designs, which rely on a static structure for all queries/tasks, \ourmethod adaptively crafts customized topologies for different domains and tasks, serving as a fully autonomous and flexible assistant for multi-agent system establishment and deployments.

Our contribution can be summarized as follows:
\vspace{-0.7em}

\begin{itemize}[leftmargin=*,itemsep=-0.3em]
\item[\ding{182}] \textbf{\textit{Protocol Proposal.}} 
We propose the \textit{first} communication protocol tailored for LLM-powered multi-agent systems, MACP, which comprehensively regulates multi-agent topology design across three dimensions: \textit{performance, adaptability, and robustness}, and incisively highlights the shortcomings of existing designs.

% \vspace{-0.3em}
\item[\ding{183}] \textbf{\textit{Practical Solution.}} 
We present \ourmethod, an effective, adaptive, and robust designer of LLM-powered multi-agent communication graphs. By leveraging a variational graph auto-encoder to construct and process the multi-agent network, \ourmethod decodes task-adaptive and high-performing agent communication, which is also equipped with strong robustness against agent-rooted adversarial attacks via dynamic topology adjustment.

% \vspace{-0.3em}
\item[\ding{184}] \textbf{\textit{Experimental Validation.}} 
Extensive experiments across six benchmarks show that \ourmethod is: \textbf{(1) high-performing}, surpassing state-of-the-art topologies by $0.20\%\sim4.10\%$ on MMLU and HumanEval; \textbf{(2) task-adaptive}, dynamically adjusting topology complexity with task awareness, outperforming state-of-the-art methods on MMLU with a cost of merely $1.5e+5$ compared to their $2.6e+6$, reducing token consumption by up to $92.24\%$; and \textbf{(3) adversarially robust}, defending against agent adversarial attacks with merely $0.3\%$ accuracy drop.

\end{itemize}
\vspace{-0.5em}

\vspace{-0.6em}

\section{Related Works}\label{sec:related}
\vspace{-0.3em}
\paragraph{LLM-agent Collaboration}
\vspace{-0.3em}
%While the academic community has widely recognized the success of single LLM-based agents in reasoning~\citep{cot,tot,got} and planning~\citep{huang2022language,sun2023pearl,ruan2023tptu}, collaboration among multiple LLM-based agents has swiftly emerged as a powerful approach for integrating the specialized capabilities of different agents, even exceeding the performance of individual LLMs~\cite{UIST2023_Agent-Simulate-Interaction, chen2023agentverse, chateval, chen2023gamegpt, cohen2023lm, hua2023war}. A basic form of collaboration is majority voting~\citep{chen2024compundLLM}, where agents operate independently. However, more effective multi-agent collaboration should construct an interconnected system and iterative topology that encourages interdependent interactions and deliberate decision-making~\citep{piatti2024cooperate, chen2024compundLLM}. Building on this insight, pioneering 
Recent research has explored various multi-agent communication topologies, including: \textbf{(1) Non-interactive}, where agents operate independently without inter-agent communication, as employed in systems like LATM~\citep{zhang2023astools} and LLM-Debate~\citep{arXiv2023_MultiAgent-Debate}; \textbf{(2) Chain}, where agents are arranged in a sequential structure, each receiving the output from its predecessor and passing information to its successor, utilized by ChatDev~\citep{software-dev}, MetaGPT~\citep{meta-gpt}, and L2MAC~\citep{holt2024l2mac}; \textbf{(3) Star}, where a central administrative agent (often referred to as a commander, teacher,  \textit{etc}.) directs subordinate agents, seen in AutoGen~\citep{autogen}, SecurityBot~\citep{yan2024depending}, and MiniGrid~\citep{zhou2023large}; \textbf{(4) Tree}, where a root agent hierarchically manages multiple child agents, as in SoA~\citep{ishibashi2024selforganize-mother}; and \textbf{(5) Graph}, encompassing  complete graphs~\citep{qian2024scaling,zhuge2024gptswarm} and random graphs~\citep{qian2024scaling}, among others.

\vspace{-0.7em}
\paragraph{Multi-agents as Graphs}
Graphs, as a fundamental data structure for organizing and representing relationships between entities~\citep{zhang2006introduction}, are widely adopted in the pre-LLM era as a powerful tool to facilitate effective communication in multi-agent reinforcement learning (MARL)~\citep{pesce2023learning,hu2024magraph,liu2022temporal}. With the rise of LLMs and the proliferation of LLM-based agents~\cite{ chen2023gamegpt, cohen2023lm, hua2023war}, researchers have similarly recognized that interactions among multiple agents can naturally be modeled from a graph-based perspective~\citep{chen2023agentverse,zhuge2024gptswarm,qian2024scaling,arXiv2023_Dynamic-LLM-Agent}. Early attempts are implicit, like ChatEval~\citep{chateval}, AutoGen~\citep{autogen}, and DSPy~\citep{khattab2023dspy}. More recent practices including ChatLLM~\citep{chatllm-network}, DyLAN~\citep{arXiv2023_Dynamic-LLM-Agent}, GPTSwarm~\citep{zhuge2024gptswarm}, and MacNet~\citep{qian2024scaling}, have explicitly represented the organization of multiple agents as a graph. 
%Specifically, both ChatLLM~\citep{chatllm-network} and DyLAN~\citep{arXiv2023_Dynamic-LLM-Agent} utilize a multilayer perception (MLP)-like layered graph, while MacNet~\citep{qian2024scaling} systematically evaluates various predefined topologies. GPTSwarm~\citep{zhuge2024gptswarm} parameterizes and optimizes the fully connected graph distribution. 
However, all these attempts, whether predefined or iteratively optimized, remain \textit{input-independent}. Consequently, they fail to be task-aware and adaptively design topologies that suit the complexity of the specific task.

\vspace{-0.4em}
\section{Formalization}\label{sec:formal}
\vspace{-0.4em}

This section establishes the notation, formalizes key concepts from a topology perspective, and formally defines our proposed multi-agent communication protocol.
% \vspace{-0.3em}
% \paragraph{\textbf{Topological Structure}} 
\vspace{-0.6em}
\subsection{Topology Structure}
\vspace{-0.4em}

We model the multi-agent system as a directed graph $\mathcal{G}=(\mathcal{V},\mathcal{E})$, where $\mathcal{V}=\{v_1,\dots,v_{N}\}$ represents the set of nodes (with $N=|\mathcal{V}|$) and $\mathcal{E}$ denotes the set of edges. Each node $v_i\in \mathcal{V}$ corresponds to an agent, formalized as:
\vspace{-0.3em}
\begin{equation}\label{eq:agent_definition}
v_i = \{\texttt{Base}_i, \texttt{Role}_i, \texttt{State}_i, \texttt{Plugin}_i\},%\;\;\texttt{Plugin}_i=\{\texttt{F}_j,\texttt{C}_j\}_{j=1}^{P} ,
% \vspace{-0.2em}
\end{equation}
where each agent $v_i$ is composed of four key elements: (1) $\texttt{Base}_i$, the language model instance powering $v_i$;
(2) $\texttt{Role}_i$, the agent's pre-assigned role or function;
(3) $\texttt{State}_i$, representing the agent's accumulated knowledge and interaction history; and (4) $\texttt{Plugin}_i$, a set of external tools or plugins available to $v_i$, such as web searchers~\citep{ma2023laser}, code compilers~\citep{autogpt,autogen,meta-gpt,bouzenia2024repairagent,ishibashi2024selforganize-mother}, or file readers~\citep{zhuge2024gptswarm,autogpt}. Each LLM-based agent $v_i$ receives  prompt $\mathcal{P}$ and generates response $\mathcal{R}_i$:
\vspace{-0.3em}
\begin{equation}
\mathcal{R}_i = v_i(\mathcal{P}) = v_i(\mathcal{P}_\text{sys},\mathcal{P}_\text{usr}),
% \vspace{-0.3em}
\end{equation}
where $\mathcal{P}_\text{sys}=\{\texttt{Role}_i,\texttt{State}_i\}$ represents the system prompt encompassing its role and state, and $\mathcal{P}_\text{usr}$ denotes the user prompt, which possibly includes the
given tasks, responses/instructions from other agents and externally retrieved knowledge.

The connectivity of $\mathcal{G}$ can also be characterized by a (non-symmetric) adjacency matrix $\mathbf{A}\in\{0,1\}^{N\times N}$, where $\mathbf{A}[i,j]=1$ if $e_{ij}=(v_i,v_j)\in\mathcal{E}$, otherwise $0$. Each edge $e_{ij} \in \mathcal{E}$ represents the flow of information from $v_i$ to $v_j$.
% \vspace{-0.3em}
% \paragraph{\textbf{Communication Pipeline}} 

\begin{figure*}[!ht]
  \centering
  \includegraphics[width=1\linewidth]{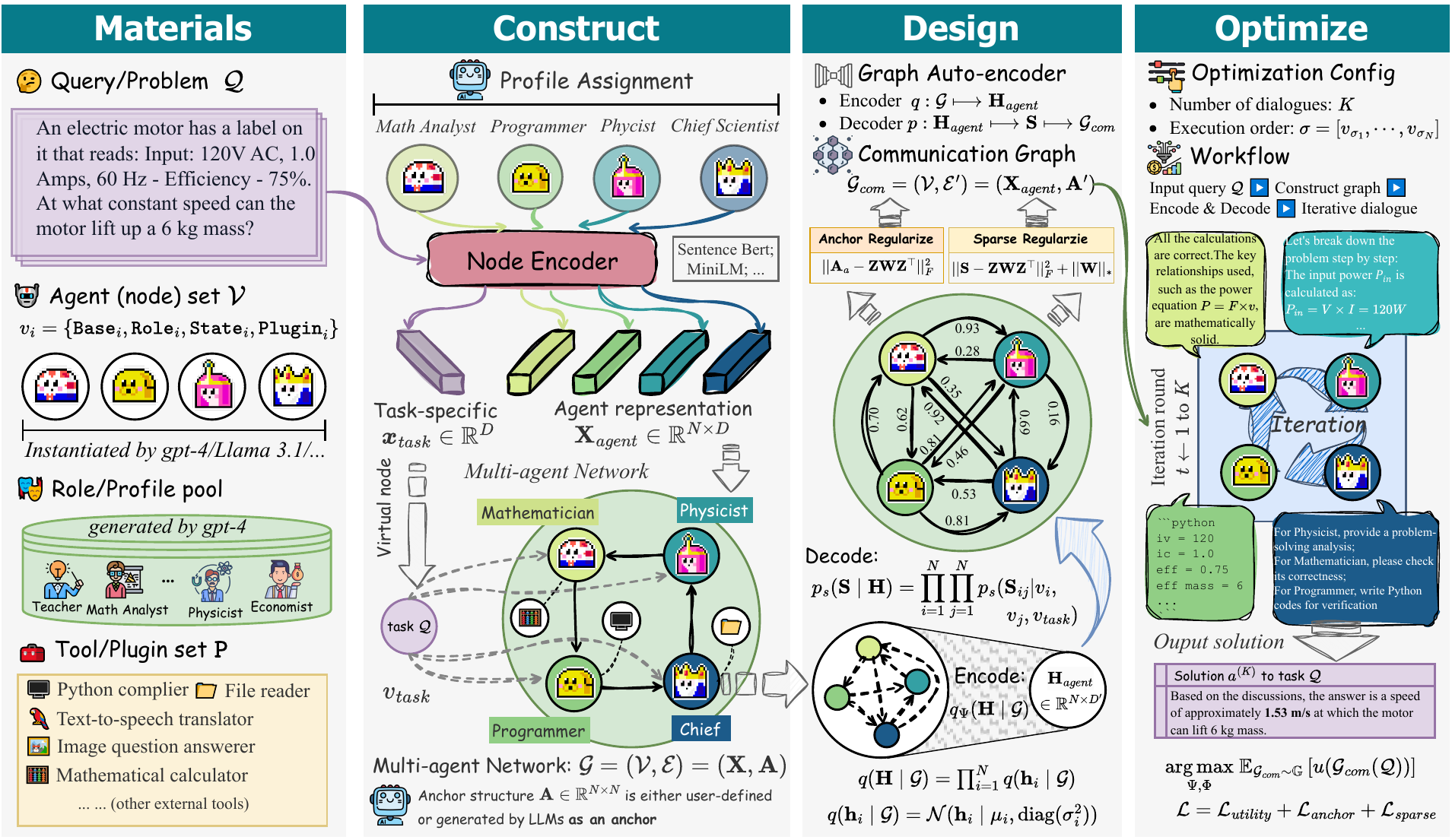}
  \vspace{-2em}
  \caption{The designing workflow of our proposed \ourmethod. }
   \label{fig:framework}
   \vspace{-1em}
\end{figure*}

\vspace{-0.4em}
\subsection{Communication Pipeline}\label{sec:communication}
\vspace{-0.4em}
Given a query/problem $\mathcal{Q}$, the multi-agent system engages in $K$ rounds of interactive utterances, which collaboratively drive the agents toward producing the final solution $a^{(K)}$ based on their cumulative dialogue exchanges.
 At the beginning of the $t$-th dialogue round, a mapping function $\phi$ is applied to determine the execution index for each agent:
\begin{equation}
\begin{aligned}
\phi:\mathcal{G}\longmapsto \sigma, \; \sigma=[v_{\sigma_1}, v_{\sigma_2}, \cdots, v_{\sigma_N}],\\\operatorname{s.t.} \forall i>j,\;\;v_{\sigma_i}\notin \mathcal{N}_\text{in}(v_{\sigma_j}),
\end{aligned}
\end{equation}
where $\sigma$ is the execution sequence of agents, $\mathcal{N}_\text{in}(v_{\sigma(j)})$ denotes the in-neighborhood of $v_{\sigma(j)}$, and the constraint ensures that an agent $v_{\sigma(i)}$ can only execute after any agent $v_{\sigma(j)}$ from which it receives information. 
Once the execution order is determined, each agent proceeds to perform input-output operations sequentially:
\begin{equation}
\mathcal{R}_i^{(t)} = v_i(\mathcal{P}^{(t)}_\text{sys}, \mathcal{P}^{(t)}_\text{usr}),\; \mathcal{P}^{(t)}_\text{usr}=\{\mathcal{Q}, \cup_{v_j \in \mathcal{N}_\text{in}(v_i)} \mathcal{R}_j^{(t)}\}
\end{equation}
where $\mathcal{R}_i^{(t)}$ represents the output of $v_i$, which could be a rationale, an answer, or a partial solution, depending on the specific context. The output $\mathcal{R}_i^{(t)}$ is generated based on the system prompt $\mathcal{P}_\text{sys}^{(t)}$ and the context prompt, consisting of the query $\mathcal{Q}$ and messages from other agents. At the end of each dialogue, an aggregation function is adopted to generate the answer/solution $a^{(t)}$:
\begin{equation}
a^{(t)} \leftarrow \operatorname{Aggregate}(\mathcal{R}_1^{(t)}, \mathcal{R}_2^{(t)},\cdots, \mathcal{R}_N^{(t)}).
\end{equation}
The implementation of the $\operatorname{Aggregate}$ function is flexible, with possible options including majority voting~\citep{chen2024compundLLM,zhuge2024gptswarm,arXiv2024_MoreAgents}, aggregating all agents' responses and delegating one agent to provide the final answer~\citep{autogen,blender,arXiv2023_Dynamic-LLM-Agent,zhang2024cut}, or simply using the output of the last agent $\mathcal{R}^{(t)}_{\sigma_N}$~\citep{qian2024scaling}. Through $K$ rounds of utterances, either predefined~\citep{qian2024scaling} or determined by an early-stopping mechanism~\citep{arXiv2023_Dynamic-LLM-Agent}, the overall system $\mathcal{G}$ produces the final answer $a^{(K)}$ for $\mathcal{Q}$.

% \vspace{-0.3em}
% \paragraph{\textbf{MACP Protocol}} 
\vspace{-0.4em}
\subsection{MACP Protocol}\label{sec:protocol}
\vspace{-0.4em}
We give the formal definition of MACP Protocol as follows:
\begin{definition}[\textbf{Multi-agent Communication Protocol}]
Given an LLM-based multi-agent system $\mathcal{G}=(\mathcal{V},\mathcal{E})$, we establish the following objective as optimization principle:
\begin{equation}\label{eq:macp}
\small
% \underset{\mathcal{G}\in\mathbb{G}}{\min} \operatorname{MACP}_{\beta} (\mathcal{G})\triangleq  \Bigl[ -u\bigl(\mathcal{G}(\mathcal{Q})\bigl) + \beta_1\cdot||\mathcal{G}|| +  \beta_2\cdot\bigl|\hat{\mathcal{G}}(\hat{\mathcal{Q}}) - \mathcal{G}(\mathcal{Q})\bigl| \Bigl],
\underset{\mathcal{G}\in\mathbb{G}}{\min}  \Bigl[ -u\bigl(\mathcal{G}(\mathcal{Q})\bigl) + \beta_1\cdot||\mathcal{G}|| +  \beta_2\cdot\bigl|\hat{\mathcal{G}}(\hat{\mathcal{Q}}) - \mathcal{G}(\mathcal{Q})\bigl| \Bigl],
\end{equation}
where $\mathbb{G}$ represents the feasible parameter space of $\mathcal{G}$, $u(\cdot)$ is the utility evaluator, $||\mathcal{G}||$ measures the computational and communication overhead of the entire graph, and $\hat{\mathcal{Q}}$ and $\hat{\mathcal{G}}$ denote the query description and the multi-agent system after adversarial perturbation, respectively. The first term in \Cref{eq:macp} corresponds to \textbf{high performance}, aiming to maximize the utility of the system's output; the second term addresses \textbf{task-adaptiveness}, seeking to minimize system complexity to reduce power consumption and economic cost; and the third term focuses on \textbf{robustness}, constraining the deviation of system output under adversarial attacks.
\end{definition}

% \textbf{Multi-agent Communication Protocol (MACP)}: \textit{For a given task/query $q$, an optimal communication topology $\mathcal{T}(q)$ for LLM-based multi-agent systems should satisfy the following criteria: (1) \emph{Efficiency}: The solution $\hat{s}(q)$ provided by the communication protocol should minimize the error $\epsilon(\hat{s}, s^) = |\hat{s}(q) - s^(q)|$, where $s^*(q)$ is the true solution. (2) \emph{Complexity-adaptiveness}: The communication structure $\mathcal{T}(q)$ should adapt to the task complexity $\mathcal{C}(q)$, seeking to minimize the communication cost $C(\mathcal{T}(q))$, subject to $C(\mathcal{T}(q)) \propto \mathcal{C}(q)$. (3) \emph{Adversarial robustness}: Under adversarial perturbations $\delta q$, the protocol should ensure robustness by bounding the solution deviation $|\hat{s}(q + \delta q) - \hat{s}(q)| \leq \epsilon_\text{adv}$, where $\epsilon_\text{adv}$ is a predefined tolerance.}

% \clearpage

\vspace{-0.4em}
\section{G-Designer}
\vspace{-0.4em}
\Cref{fig:framework} illustrates how \ourmethod adaptively designs communication topologies for any given query. Specifically, the process begins with a few ``raw materials'': the input query $\mathcal{Q}$, the agent set $\mathcal{V}$, the profile pool, and the toolset. In the \textit{Construct} stage, \ourmethod leverages a node encoder to construct a multi-agent network along with a task-specific virtual node. In the \textit{Design} stage, a graph auto-encoder is employed to decode the communication graph topology $\mathcal{G}_{com}$, which is leveraged for multi-round inter-agent collaboration in the \textit{Optimize} stage. 

\vspace{-0.4em}
\subsection{Multi-agent Network Construction}\label{sec:multi-agent-network}
\vspace{-0.4em}
Given an input query $\mathcal{Q}$ and a set of LLM-agents $\mathcal{V}$, \ourmethod aims to design a task-adaptive and effective communication topology $\mathcal{G}_{com}$. We begin by assigning each agent a unique role and profile, as previous research~\citep{multi-persona} has shown that assigning distinct personas or roles to LLM-based agents can enhance cognitive synergy. Based on these roles, different external tools are allocated to the agents (\textit{e.g.}, Mathematica for a math analyst, Python compiler for a programmer). Thus, we successfully initialize each agent $v_i$ as $\{\texttt{Base}_i, \texttt{Role}_i, \texttt{State}_i, \texttt{Plugin}_i\}$, as defined in \Cref{eq:agent_definition}.

We proceed to construct a structured multi-agent network as input to \ourmethod, represented as $\mathcal{G}=(\mathbf{X}_{agent},\mathbf{A})$, where $\mathbf{X}_{agent}\in\mathbb{R}^{N\times D}$ is the node (agent) feature matrix and $\mathbf{A}\in\mathbb{R}^{N\times N}$ represents the connectivity matrix. For the feature matrix $\mathbf{X}_{agent}$, we employ a node encoder to transform each agent’s unique profile into a fixed-length embedding representation:
\begin{equation}\label{eq:node-encoder}
\vspace{-0.4em}
\small
\mathbf{x}_i \leftarrow \operatorname{NodeEncoder}\left(\mathcal{T}(\texttt{Base}_i), \texttt{Role}_i, \mathcal{T}(\texttt{Plugin}_i)\right),
\end{equation}
where $\mathcal{T}(\cdot)$ extracts the textual description of the agent’s LLM backbone and its assigned plugins, and $\operatorname{NodeEncoder}$ can be realized using small and lightweight text embedding models~\citep{reimers2019sentence}. After encoding the individual agents, we aim to ensure that the multi-agent network incorporates information related to the query $\mathcal{Q}$, as this query-dependent approach enables \ourmethod to be task-aware and adaptive. To this end, we introduce an additional \textit{task-specific virtual global node} $v_{task}$, which is bidirectionally connected to all agent nodes, enabling a global "storage sink" and facilitating smoother information flow among agents~\citep{shirzad2023exphormer,tan2023virtual,rosenbluth2024distinguished}. This task node is encoded by the $\operatorname{NodeEncoder}$ as follows: $\mathbf{x}_{task} \leftarrow \operatorname{NodeEncoder}(\mathcal{Q})$.

After obtaining the agent node features $\mathbf{X}_{agent} = [\mathbf{x}_1, \mathbf{x}_2, \dots, \mathbf{x}_N]^\top$ and the task-specific embedding $\mathbf{x}_{task}$, we provide a simple \textit{anchor topology} $\mathbf{A}_{anchor} \in \{0,1\}^{N \times N}$, which serves as a starting point for \ourmethod's topology design process. For instance, given a code generation task with three agents: manager/programmer/code reviewer, the anchor topology could be configured as a chain structure, \textit{i.e.}, ``manager $\rightarrow$ programmer $\rightarrow$ reviewer'', reflecting the typical workflow of code completion. The anchor topology, being either user-defined or automatically generated by LLMs, is often simple and sub-optimal\footnote{We discuss the substantial performance improvement of \ourmethod over the anchor topology in \Cref{sec:analysis}.}. However, it provides a foundational reference and prior knowledge for \ourmethod's subsequent optimization process. We incorporate the task-specific vertex $v_{task}$ and its corresponding edges and obtain $\Tilde{\mathbf{A}}_{anchor}\in\{0,1\}^{(N+1)\times(N+1)}$. Consequently, we establish a task-specific multi-agent network $\Tilde{\mathcal{G}}$:
\begin{equation}\label{eq:task_graph}
\begin{aligned}
\vspace{-0.4em}
\Tilde{\mathcal{G}} & = \bigl(\begin{bmatrix}
   \mathbf{X}_{agent} \\
   \mathbf{x}_{task}^\top
   \end{bmatrix}, \Tilde{\mathbf{A}}_{anchor}\bigl)=(\Tilde{\mathcal{V}},\Tilde{\mathcal{E}})\\
   &= \bigl(\mathcal{V}\cup \{v_{task}\},\mathcal{E}\cup \{\overset{\longleftrightarrow}{(v_i,v_{task})}|v_i\in\mathcal{V})\}\bigl),
   \end{aligned}
\end{equation}
where $\scriptsize \begin{bmatrix}
   \mathbf{X}_{agent} \\
   \mathbf{x}_{task}^\top
   \end{bmatrix}$ can also be jointly denoted as $\Tilde{\mathbf{X}}$.

\vspace{-0.4em}
\subsection{Designing Communication Topology}
\vspace{-0.4em}
Building upon the task-specific multi-agent network $\Tilde{\mathcal{G}}$, \ourmethod seeks to establish a more fine-grained and precise communication topology $\mathcal{G}_{com}$. Drawing inspiration from the variational graph auto-encoder (VGAE) framework~\citep{kipf2016variational,zhao2024causality}, \ourmethod employs a VGAE-based encoder-decoder $f_v$ to generate the multi-agent interaction topology:
\begin{equation}
\small
\mathcal{G}_{com} = f_v(\Tilde{\mathcal{G}};\Theta_v) = p(\mathcal{G}_{com}\;|\;\mathbf{H})q(\mathbf{H}\;|\;\Tilde{\mathbf{X}},\Tilde{\mathbf{A}}_{anchor}),
\end{equation}
where $f_v$ is the encoder-decoder architecture with parameters $\Theta_v$, $q(\cdot)$ is the encoder module, $p(\cdot)$ is the decoder module. The encoder utilizes posterior probabilities to encode the node embeddings
into low-dimensional latent vector representations $\mathbf{H}_{agent}$, which can be formulated as:
\begin{equation}
\begin{aligned}
q(\mathbf{H}_{agent}\;|\;\Tilde{\mathbf{X}},\Tilde{\mathbf{A}}_{anchor})=\prod_{i=1}^{N}q(\mathbf{h}_i\;|\;\Tilde{\mathbf{X}},\Tilde{\mathbf{A}}_{anchor}),\\
q(\mathbf{h}_i\;|\;\Tilde{\mathbf{X}},\Tilde{\mathbf{A}}_{anchor}) = \mathcal{N}(\mathbf{h}_i\;|\;\boldsymbol{\mu}_i,\operatorname{diag}(\boldsymbol{\sigma}_i^2)),
\end{aligned}
\end{equation}
where $\mu = \operatorname{GNN}_\mu(\Tilde{\mathbf{X}},\Tilde{\mathbf{A}}_{anchor};\Theta_\mu)$ is the matrix of mean vectors $\mu_i$; similarly $\log(\sigma) = \operatorname{GNN}_\sigma(\Tilde{\mathbf{X}},\Tilde{\mathbf{A}}_{anchor};\Theta_\sigma)$. The choice of GNN backbone can be customized as needed; here, we utilize a simple two-layer GCN~\citep{kipf2016semi}. $\mathbf{h}_i$, $\boldsymbol{\mu}_i$, and $\boldsymbol{\sigma}_i$ denote the $i$-th column of $\mathbf{H}$, $\boldsymbol{\mu}$, and $\boldsymbol{\sigma}$, respectively. The encoder $q(\cdot)$ is parameterized by $\Theta_e=\{\Theta_\mu,\Theta_\sigma\}$. Following the encoding phase, the decoder employs the latent representations to generate a comprehensive blueprint for multi-agent communication. More specifically, the decoder $q(\cdot)=q_c\circ q_s$ first constructs a parameterized, sketched graph $\mathbf{S}$, which is then refined into the final multi-agent communication topology:
\begin{equation}
p(\mathcal{G}_{com} \;|\; \mathbf{H}_{agent}) = \int_{\mathbf{S}} p_{c}(\mathcal{G}_{com} \;|\; \mathbf{S}) p_s(\mathbf{S} \;|\; \mathbf{H}_{agent}) \;d\mathbf{S}.
\end{equation}
At the first step,  $p_s(\cdot)$ constructs the sketched adjacency matrix $\mathbf{S}$ from the latent representation $\mathbf{H}_{agent}$:
\begin{equation}
\small
p_s(\mathbf{S} \;|\; \mathbf{H}_{agent}) = \prod_{i=1}^N\prod_{j=1}^N p_s(\mathbf{S}_{ij}\;|\;\mathbf{h}_i,\mathbf{h}_j,\mathbf{h}_{task};\Theta_d),
\end{equation}
whose detailed derivation is as follows:
\begin{equation}\label{eq:s_ij}
\begin{aligned}
p_s(\mathbf{S}_{ij}=1\;|\;\mathbf{h}_i,\mathbf{h}_j,\mathbf{h}_{task})=g(\mathbf{h}_i,\mathbf{h}_j,\mathbf{h}_{task}),\\
=\operatorname{Sigmoid}((\log(\epsilon) - \log(1-\epsilon) + \varpi_{ij})/\tau),
\end{aligned}
\end{equation}
where $\varpi=\operatorname{FFN}_d([\mathbf{h}_i,\mathbf{h}_j,\mathbf{h}_{task}])$ with $\operatorname{FFN}_d$ parameterized by $\Theta_d$, $\epsilon \sim \operatorname{Uniform}(0,1)$, and $\tau$ denotes the temperature coefficient. When $\tau$ approaches zero, \Cref{eq:s_ij} essentially return the Bernouli sampling result for $\mathbf{S}_{ij}$. 
The resulting matrix $\mathbf{S}\in[0,1]^{N\times N}$ represents a densely-connected, non-negative graph distribution, indicating an overly complex and resource-intensive pair-wise communication structure, which is not yet suitable for guiding multi-agent collaboration. To align with \ourmethod's objectives of task adaptiveness and minimizing costs, we apply a refinement decoder $p_c(\cdot)$ to refine the sketched $\mathbf{S}$ into a compact, sparse, and highly informative communication graph, instantiated by a regularization objective:
\begin{equation}\label{eq:low_rank}
\begin{aligned}
p_c:\; \underset{\Tilde{\mathbf{S}}\in\mathbb{S}}{\arg \max}\; 1/2||\mathbf{S} - \mathbf{Z}\mathbf{W}\mathbf{Z}^\top||^2_F + \zeta||\mathbf{W}||_* +\\
1/2||\mathbf{A}_{anchor} - \mathbf{Z}\mathbf{W}\mathbf{Z}^\top||^2_F,\; \operatorname{s.t.}\; \Tilde{\mathbf{S}}=\mathbf{Z}\mathbf{W}\mathbf{Z}^\top,
\end{aligned}
\end{equation}
where $\mathbf{Z}\in\mathbb{R}^{N\times r}$  is the top-$r$ columns of left singular matrix $\mathbf{S}$, $\zeta$ is a coefficient hyperparameter, $\mathbf{W}\in\mathbb{R}^{r\times r}$ is an optimizable weight matrix, $||\cdot||_F$ denotes the Frobenius norm and $||\mathbf{W}||_*=\sum_i\lambda_i$ where $\lambda_i$ is the $i$-th singular value of $\mathbf{W}$.
$\Tilde{\mathbf{S}}\in\mathbb{R}^{N\times N}$ is the desired sparse topology, which is decomposed as $\mathbf{Z}\mathbf{W}\mathbf{Z}^\top$. In \Cref{eq:low_rank}, the first and second terms are jointly denoted as \textit{anchor regularization}, which encourage the learned $\Tilde{\mathbf{S}}$ to maintain similarity with both the original $\mathbf{S}$ and the anchor topology. The third term, denoted as \textit{sparsity regularization}, though appearing to minimize the nuclear norm of $\mathbf{W}$, essentially sparsifies $\Tilde{\mathbf{S}}$, since $||\Tilde{\mathbf{S}}||_*=||\mathbf{W}||_*$ holds due to $\mathbf{Z}^\top\mathbf{Z}=\mathbbm{I}_{r\times r}$. Therefore, \Cref{eq:low_rank} achieves two key goals: (1) producing a sparse, refined communication topology, and (2) constraining the design to remain grounded in practical intuition. The resulting communication can be represented as follows:
\begin{equation}
\small
\mathcal{G}_{com} = (\mathcal{V}, \mathcal{E}_{com}), \mathcal{E}_{com} = \{(i,j)\;|\;\Tilde{\mathbf{S}}_{ij}\neq 0 \wedge (i,j)\in\mathcal{E}\}).
\end{equation}
At this stage, we have successfully distilled a lightweight and informative collaboration network $\mathcal{G}_{com}$ from the sketched task-specific network $\Tilde{\mathcal{G}}$, which is now ready to guide inter-agent message passing in the following process.

\begin{table*}[!t]
\centering
\caption{Performance comparison with three types of baselines, including single-agent execution, spatial communication, and temporal communication. The best results are in bold, and the runner-ups are underlined.  {All methods, except for the single-agent category, utilize \textbf{five} \llmname{gpt-4}-based agents.} ``Mul.'', ``Ada.'', and ``Rob.'' indicate whether the method supports a multi-agent setting, whether it is task-adaptive, and whether it is adversarially robust, respectively. \textcolor{darksalmon}{\XSolidBrush}, {\large\textcolor{Dandelion}{{\ding{51}}{\small{\kern-0.7em\ding{55}}}}} and \textcolor{green(pigment)}{\Checkmark} signifies no/partial/full support in these aspects.}
\vspace{-0.1em}
\label{tab:rq1_performance}
\renewcommand\tabcolsep{5.3pt}
\renewcommand\arraystretch{1.1}

\resizebox{\linewidth}{!}{
\begin{tabular}{l|ccc|ccccccc}
\Xhline{1.2pt}
\rowcolor{CadetBlue!20} 
{\textbf{Method}} & \textbf{Mul.} & \textbf{Ada.} & \textbf{Rob.} & \textbf{MMLU} & \textbf{GSM8K} & \textbf{MultiArith} & \textbf{SVAMP} & \textbf{AQuA} & \textbf{HumanEval} & {\textbf{Avg.}} \\
\Xhline{1.2pt}
Vanilla & \textcolor{darksalmon}{\XSolidBrush} & \textcolor{darksalmon}{\XSolidBrush} & \textcolor{darksalmon}{\XSolidBrush} & 82.14 & 85.40 & 93.15 & 87.18 & 70.34 & 71.68 & 81.65\\
\hline

\rowcolor{gray!10}CoT & \textcolor{darksalmon}{\XSolidBrush}  & \textcolor{darksalmon}{\XSolidBrush} & \textcolor{darksalmon}{\XSolidBrush} & 82.65\red{0.51} & 87.17\red{1.77} & 94.79\red{1.64} & 88.32\red{1.14} & 73.91\red{3.57} & 75.52\red{3.84} & 83.73\\

ComplexCoT & \textcolor{darksalmon}{\XSolidBrush}  & \textcolor{darksalmon}{\XSolidBrush} & \textcolor{darksalmon}{\XSolidBrush} & 83.78\red{1.64} & 87.62\red{2.22} & 95.86\red{2.71} & 90.17\red{2.99} & 77.58\red{7.24} & 74.94\red{3.26} & 84.99\\

\rowcolor{gray!10}SC (CoT) & \textcolor{darksalmon}{\XSolidBrush}  & \textcolor{darksalmon}{\XSolidBrush} & \textcolor{darksalmon}{\XSolidBrush} & 82.66\red{0.52} & 87.93\red{2.53} & 96.88\red{3.73} & 88.69\red{1.51} & 75.08\red{4.74} & 77.30\red{5.62} & 84.75 \\

SC (ComplexCoT) & \textcolor{darksalmon}{\XSolidBrush}  & \textcolor{darksalmon}{\XSolidBrush} & \textcolor{darksalmon}{\XSolidBrush} & 83.65\red{1.51} & 86.14\blue{0.74} & 96.94\red{3.79} & 89.72\red{2.54} & 77.69\red{7.35} & 77.94\red{6.26} & 85.35\\

% !!AutoGPT & \textcolor{darksalmon}{\XSolidBrush}   & \textcolor{darksalmon}{\XSolidBrush} & \textcolor{darksalmon}{\XSolidBrush} & 83.65\red{1.51} & 86.14\blue{0.74} & 96.94\red{3.79} & 89.72\red{2.54} & 77.69\red{7.35} & 77.94\red{6.26} & 85.35\\

\rowcolor{gray!10}PHP & \textcolor{green(pigment)}{\Checkmark} & \textcolor{darksalmon}{\XSolidBrush} &  \textcolor{darksalmon}{\XSolidBrush}  & 83.45\red{1.31} & \textbf{95.50}\red{10.1} & \underline{98.10}\red{2.84} & 90.02\red{3.44} & {79.00}\red{8.66} & 82.96\red{11.36} & 88.17\\

\hline

Chain& \textcolor{green(pigment)}{\Checkmark} &  \textcolor{darksalmon}{\XSolidBrush}  & \textcolor{darksalmon}{\XSolidBrush} & 82.35\red{0.21} & 85.57\red{0.17} & 94.38\red{1.23} & 83.41\blue{3.77} & 70.94\red{0.60} & 80.88\red{9.20} & 82.92\\

\rowcolor{gray!10}Star & \textcolor{green(pigment)}{\Checkmark}&  \textcolor{darksalmon}{\XSolidBrush}  & \textcolor{darksalmon}{\XSolidBrush} & 80.79\blue{1.35} & 85.55\red{0.15} & 93.79\blue{0.64} & 88.09\red{0.91} & 68.57\blue{1.77} & 75.65\red{3.97} & 82.07\\

Tree& \textcolor{green(pigment)}{\Checkmark} &  \textcolor{darksalmon}{\XSolidBrush}  & \textcolor{darksalmon}{\XSolidBrush} & 81.89\blue{0.25} & 84.56\blue{0.84} & 94.60\red{1.45} & 89.25\red{2.07} & 72.84\red{2.50} & 77.38\red{5.70}& 83.42 \\

\rowcolor{gray!10}Complete Graph & \textcolor{green(pigment)}{\Checkmark} &  \textcolor{darksalmon}{\XSolidBrush} & \textcolor{darksalmon}{\XSolidBrush} & 83.15\red{1.01} & 86.49\red{1.09} & 97.20\red{4.05} & 89.48\red{2.30} & \underline{79.21}\red{8.87} & 83.75\red{12.07} & 86.55\\

Random Graph & \textcolor{green(pigment)}{\Checkmark} &  \textcolor{darksalmon}{\XSolidBrush}  & \textcolor{darksalmon}{\XSolidBrush} & 83.76\red{1.62} & 86.14\red{0.74} & 95.46\red{2.31} & 85.41\blue{1.77} & 74.07\red{3.73} & 82.66\red{10.98}& 84.58 \\

\rowcolor{gray!10}AutoGen & \textcolor{green(pigment)}{\Checkmark}  &  \textcolor{darksalmon}{\XSolidBrush}  & \textcolor{darksalmon}{\XSolidBrush} & 82.13\blue{0.01} & 90.06\red{7.92} & 93.80\red{0.65} & 88.44\blue{1.26} & 73.65\red{3.31} & 85.41\red{13.73}& 85.58 \\

MetaGPT & \textcolor{green(pigment)}{\Checkmark}  & \textcolor{darksalmon}{\XSolidBrush}  & \textcolor{darksalmon}{\XSolidBrush}  & - & - & - & - & - & 85.90\red{14.22}& 84.90 \\

\rowcolor{gray!10}LLM-Blender & \textcolor{green(pigment)}{\Checkmark}  &  \textcolor{darksalmon}{\XSolidBrush}  &  \textcolor{darksalmon}{\XSolidBrush} & 81.22\blue{0.92} & 89.17\red{3.77} & 94.27\red{1.12} & 88.77\red{1.59} & 77.05\red{6.71} & - & 86.09 \\

LLM-Debate & \textcolor{green(pigment)}{\Checkmark} & \textcolor{darksalmon}{\XSolidBrush} & \textcolor{green(pigment)}{\Checkmark} & 83.69\red{1.55} & 90.23\red{4.83} & 96.27\red{3.12} & 90.56\red{3.38} & 77.52\red{7.18} & 83.79\red{12.11} & 87.01 \\

\rowcolor{gray!10}DyLAN & \textcolor{green(pigment)}{\Checkmark} & \textcolor{Dandelion}{{\ding{51}}{\small{\kern-0.7em\ding{55}}}} & \textcolor{green(pigment)}{\Checkmark} & 80.16\blue{1.98} & 88.16\red{2.76} & 94.27\red{1.12} & 87.40\red{0.22} & 74.16\red{3.82} & \underline{89.70}\red{18.02} & 85.64 \\

GPTSwarm & \textcolor{green(pigment)}{\Checkmark} & \textcolor{Dandelion}{{\ding{51}}{\small{\kern-0.7em\ding{55}}}} & \textcolor{green(pigment)}{\Checkmark} & \underline{83.98}\red{1.84} & {89.74}\red{4.34} & {97.84}\red{4.69} & 86.42\blue{0.76} & 78.16\red{7.82} & 88.49\red{16.81} & 87.32 \\ 

\hline

\rowcolor{gray!10}\ourmethod & \textcolor{green(pigment)}{\Checkmark}  & \textcolor{green(pigment)}{\Checkmark} & \textcolor{green(pigment)}{\Checkmark} & \textbf{84.50}\red{2.36} & \underline{95.07}\red{9.67} & \textbf{{98.30}\red{5.15}} & \textbf{91.85}\red{4.67} & \textbf{79.47}\red{9.13} & \textbf{89.90}\red{18.22} & \textbf{89.84}\\

\Xhline{1.2pt}
\end{tabular}
}
\vspace{-1.3em}
\end{table*}

\vspace{-0.4em}
\subsection{Optimizing G-Designer}
\vspace{-0.4em}
Upon obtaining $\mathcal{G}_{com}$, the multi-agent utterances and dialogues can proceed as usual using $\mathcal{G}_{com}$, as detailed in \Cref{sec:communication}. After $K$ rounds of interaction, the agents converge to a final solution $a^{(K)}=\mathcal{G}_{com}(\mathcal{Q})$. We then give the following optimization objective:
\begin{equation}\label{eq:utility}
\underset{\Theta_e,\Theta_d}{\arg\min}\; \mathbb{E}_{\Theta_e,\Theta_d \sim \Omega} \Bigl[u\bigl(\mathcal{G}_{com}(\mathcal{Q})\bigl) \Bigl],
\end{equation}
where $\Theta_e$ and $\Theta_d$ are the parameters of the encoder $q(\cdot)$ and decoder $p(\cdot)$, respectively, $\Omega$ is the parameter space and $\mathbb{E}(\cdot)$ denotes the mathematical expectation. \Cref{eq:utility} aims to maximize the utility of the generated solution, but it is inherently intractable and non-differentiable, as $u(\cdot)$ often depends on external API calls~\citep{li2023api,mmlu}. To address this, following standard approaches in multi-agent structure design~\citep{zhuge2024gptswarm,zhang2024cut}, we apply policy gradient~\citep{williams1992simple} to approximate and optimize \Cref{eq:utility}:
\begin{equation}
\small
\nabla_{\Theta} \mathbb{E}_{\Theta \sim \Omega}\Bigl[u\bigl(\mathcal{G}_{com}(\mathcal{Q})\bigl) \Bigl]\approx \frac{1}{M} \sum_{k= 1}^M u(a_m^{(K)}) \nabla_\Theta(P(\mathcal{G}_{k})),
\end{equation}
where $\mathbf{\Theta} = \{\Theta_e,\Theta_d\}$, $\{\mathcal{G}_{k}\}_{m=1}^M$ are indepently samples from $\mathcal{G}_{com}$, and $\{a_m^{(K)}\}_{m=1}^M$ are the corresponding output. $P(\mathcal{G}_{k})$ calculates the probability of $\mathcal{G}_{k}$ being sampled, which can be expressed as $P(\mathcal{G}_{k}) = \prod_{i=1}^N\prod_{j=1}^N \Tilde{\mathbf{S}}_{ij}$. Through iterative optimization guided by \Cref{eq:low_rank,eq:utility} over a limited set of queries as the ``training set'', \ourmethod efficiently develops task-awareness and the capability to strategically design the agent network, achieving truly task-customized multi-agent topology design.

\vspace{-0.5em}
\paragraph{Optimization configuration} The overall training objective of our method is formulated as $\mathcal{L}_{\ourmethod}=\mathcal{L}_{utility}+\mathcal{L}_{anchor}+\mathcal{L}_{sparse}$, where $\mathcal{L}_{utility}$ represents the optimization target from \Cref{eq:utility}, $\mathcal{L}_{anchor}$ corresponds to the first and third terms in \Cref{eq:low_rank}, and $\mathcal{L}_{sparse}$ is the second term. Given a benchmark $\{\mathcal{Q}_i\}_{i=1}^D$ consisting of $B$ queries, \ourmethod begins by optimizing with a small subset of $B'$ queries and fixes the learned parameters for testing on the remaining $(B-B')$ queries. The whole algorithm workflow of \ourmethod is depicted in Algorithm~\ref{alg:algo}.

% \clearpage
\vspace{-0.5em}
\section{Experiments}

% In this section, we conduct extensive experiments to answer the following research questions: 
% \vspace{-0.8em}
% \begin{itemize}[leftmargin=*,itemsep=-0.4em]
% \item (\textbf{RQ1}) Can \ourmethod design \textit{effective} and \textit{high-performing} multi-agent communication topologies?
% % \vspace{-0.6em}
% \item (\textbf{RQ2}) Can \ourmethod generate more \textit{task-adaptive} topologies, resulting in less token consumption?
% % \vspace{-0.6em}
% \item (\textbf{RQ3}) Is \ourmethod adversarially \textit{robust}?
% % \vspace{-0.6em}
% \item (\textbf{RQ4}) How sensitive is the proposed \ourmethod sensitive to its key components and parameters?
% \end{itemize}
% \vspace{-0.6em}

\vspace{-0.5em}
\subsection{Experimental Setup}
\vspace{-0.4em}
\paragraph{Datasets and Metrics}
We evaluate \ourmethod on three categories of datasets:
$\blacksquare$ \textbf{General Reasoning}: MMLU~\citep{mmlu};
$\blacksquare$ \textbf{Mathematical Reasoning:} GSM8K~\citep{arXiv2021_Verifier-Math}, MultiArith~\citep{roy2016solving}, SVAMP~\citep{patel2021nlp}, and AQuA~\citep{ling2017program}; $\blacksquare$ \textbf{Code:}  HumanEval~\citep{human-eval}. We include the dataset statistics in \Cref{tab:dataset_stats}.

\vspace{-1em}
\paragraph{Baselines} For single-agent approaches, we select \textbf{COT}~\citep{cot}, \textbf{ComplexCoT}~\citep{fu2022complexity}, \textbf{Self-Consistency}~\citep{wang2023selfconsistency}, and \textbf{PHP}~\citep{PHPrompting}.
For multi-agent topologies, we select \textbf{Chain}, \textbf{Star}, and \textbf{Tree} (formally defined in~\citep{qian2024scaling}), \textbf{Complete Graph} and \textbf{Random Graph} , \textbf{AutoGen}~\citep{autogen}, \textbf{MetaGPT}~\citep{meta-gpt}, \textbf{LLM-Debate}~\citep{arXiv2023_MultiAgent-Debate}, \textbf{LLM-Blender}~\citep{blender}, \textbf{DyLAN}~\citep{arXiv2023_Dynamic-LLM-Agent}, and \textbf{GPTSwarm}~\citep{zhuge2024gptswarm}.

\vspace{-1em}
\paragraph{Implementation Details} 
We access the GPT via the OpenAI API, and mainly test on \texttt{gpt-4-1106-preview} (\texttt{gpt-4}) and \texttt{gpt-3.5-turbo-0125} (\texttt{gpt-3.5}). We set \texttt{temperature} to 0 for the single execution and single agent baselines and 1 for multi-agent methods. We set a summarizer agent to aggregate the dialogue history and produce the final solution $a^{(K)}$, with $K=3$ across all experiments. The $\operatorname{NodeEncoder}(\cdot)$ is implemented using \texttt{all-MiniLM-L6-v2}~\citep{wang2020minilm}, with the embedding dimension set to $D=384$. The anchor topology $\mathbf{A}_{anchor}$ is predefined as a simple chain structure. The sampling times $M$ are set as $10$, and $\tau=1e-2$ and $\zeta=1e-1$ are set for all experiments. We provide explicit agent profiling for multi-agent methods, following the classical configurations in LLM-MA systems~\citep{arXiv2023_Dynamic-LLM-Agent,zhuge2024gptswarm,yin2023exchange}, and use \texttt{gpt-4} to generate agent profile pools. For all benchmarks, we merely use $B'\in\{40,80\}$ queries for optimization.

\vspace{-0.5em}
\subsection{Main Results}
\vspace{-0.4em}
In this section, we conduct extensive experiments across six benchmarks to verify that \ourmethod is:
\vspace{-1em}
\paragraph{High-performing} The experimental results from \Cref{tab:rq1_performance} demonstrate that \ourmethod is effective in designing powerful LLM-MA topologies. Concretely, \ourmethod achieves the best performance in five out of six benchmarks, and on GSM8K, it trails only PHP with a $9.67\%\uparrow$ accuracy improvement. On the HumanEval benchmark, \ourmethod surpasses MetaGPT, a specialized multi-agent code generation framework, by $4.0\%$ at \textit{pass@1}, and outperforms state-of-the-art multi-agent collaboration frameworks like GPTSwarm and DyLAN by margins of $0.20\%\sim1.41\%$. 
% For general reasoning on MMLU, \ourmethod also achieves a $2.36\%\uparrow$ accuracy improvement. The performance gains on MMLU are relatively marginal, as observed in previous studies~\citep{zhuge2024gptswarm,qian2024scaling}. 
% Overall, \ourmethod demonstrates exceptional performance in topology design across a wide range of tasks.
\vspace{-1em}
\paragraph{Task-adaptive} 
%Task-aware multi-agent network design not only improves task performance but also regulates the complexity of the topology according to the task’s difficulty. 
\Cref{fig:case_study} visualizes the different topologies designed by \ourmethod for varying query difficulties on HumanEval and GSM8K. As shown in \Cref{fig:case_study}, the multi-agent topologies generated by \ourmethod are highly dependent on the specific task context and its difficulty. In \textit{Case a}, despite having five \texttt{gpt-4} agents available as design resources, \ourmethod identified the task of designing a \texttt{strlen(string)} function as relatively simple. It streamlined the topology by removing unnecessary agents and retained only a minimal ``Algorithm Designer $\rightarrow$ Programmer'' structure to solve the problem. In contrast, for the more complex \textit{Case c} and \textit{Case e}, \ourmethod crafted a more intricate communication graph. These cases highlight the strong task-adaptiveness of \ourmethod.

\vspace{-1.2em}
\paragraph{Scalable} To evaluate the scalability of \ourmethod to a larger number of agents, we report its performance across $5\sim20$ agents, as presented in \Cref{tab:agent_scale}. Notably, \ourmethod exhibits a steeper performance gain than GPTSwarm as the agent count increases. More importantly, while the complete graph and GPTSwarm incur an overwhelming token cost at $20$ agents ($5.6\sim30.3$M tokens), \ourmethod achieves superior performance with merely $6.11\%$ of GPTSwarm’s prompt token consumption, surpassing it by $2.44\%\uparrow$. These results decisively demonstrate the scalability and potential of \ourmethod\ in advancing large-scale autonomous multi-agent systems.

\begin{figure}[!t]
  \centering
  \includegraphics[width=1\linewidth]{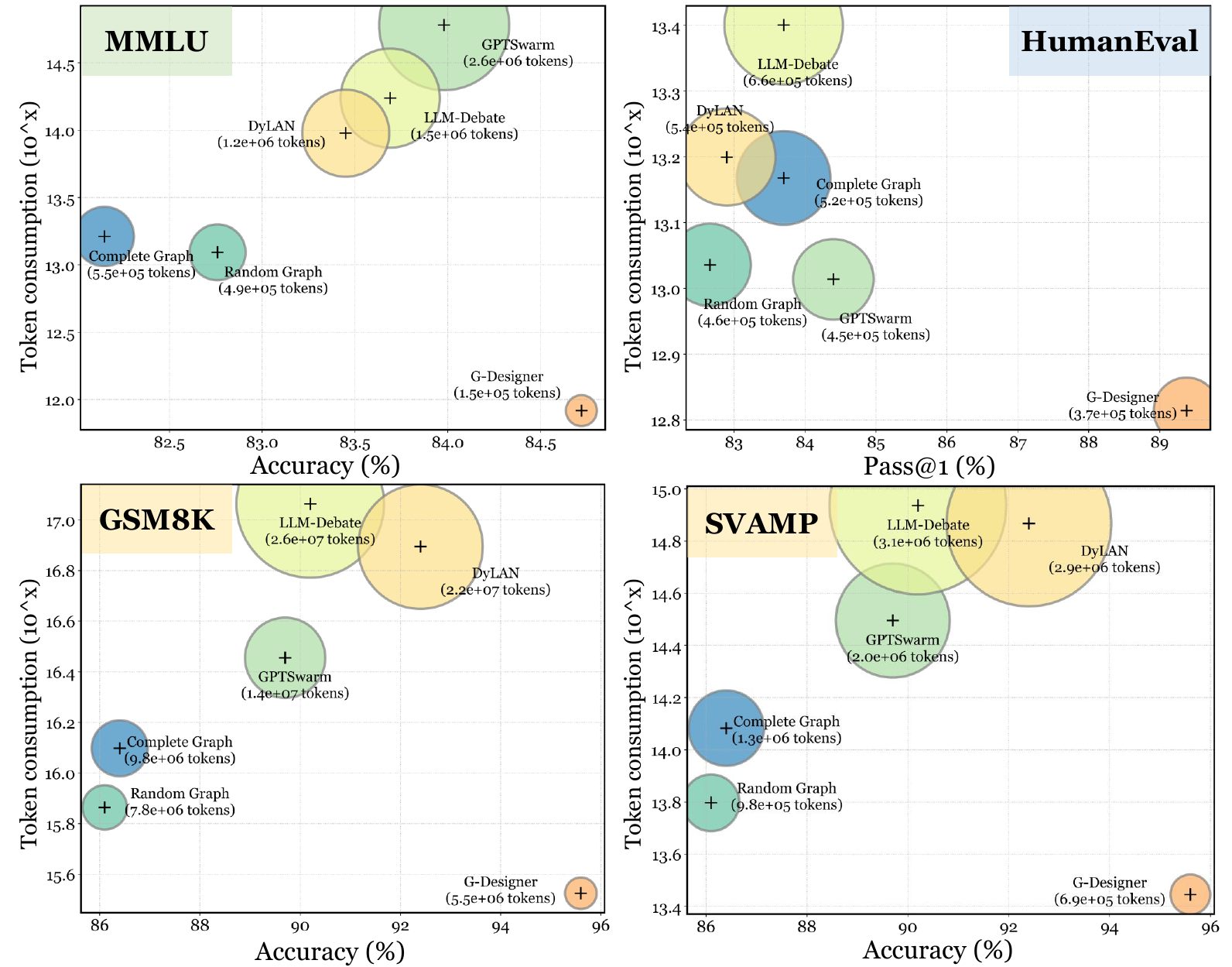}
  \vspace{-2.3em}
  \caption{Visualization of the performance metrics and prompt token consumption of different multi-agent communication topologies across
MMLU, HumanEval, GSM8K, and SVAMP. The diameter of each point is proportional to its $y$-axis value.}
   \label{fig:token}
   \vspace{-1.9em}
\end{figure}

\vspace{-1.6em}
\paragraph{Token-economical (\textit{inference})} A key benefit of \ourmethod's adaptivity is that it prevents the use of overly complex structures for simple tasks, thus minimizing unnecessary communication costs—in the case of LLM-MA, reducing token consumption. \Cref{fig:token} It illustrates the differences in prompt token consumption between \ourmethod and several representative multi-agent designs. We observe that simpler topologies, such as complete graphs and random graphs, consume fewer tokens but show significantly weaker performance. More complex communication structures, like GPTSwarm and DyLAN, achieve superior performance, albeit at the cost of excessive token consumption. For instance, DyLAN's cost on GSM8K is $2.82\times$ that of the random graph, reaching a substantial $2.2e+7$. In contrast, \ourmethod elegantly balances both efficiency and task performance, achieving the highest performance across all four benchmarks while maintaining the lowest token cost. For example, on SVAMP, \ourmethod surpasses DyLAN by $4\%$ while using only $23.7\%$ of DyLAN's token cost.

\vspace{-0.8em}
\paragraph{Resource-efficient~(\textit{training})} 
We validate \textbf{\ourmethod's training process is resource-friendly} from three dimensions: GPU cost, token cost, and wall-clock time. \Cref{tab:gpu} showcases that training \ourmethod with up to 1000 agents requires less than 4GB of memory.  \Cref{tab:resource} unveils that \ourmethod not only attains the highest accuracy but also exhibits superior token efficiency and reduced wall-clock time compared to existing baselines, underscoring its effectiveness in multi-agent collaboration.

% \begin{table}[h]
%     \centering
% \vspace{-1em}
%     \caption{Efficiency analysis. We compare the training wall-clock time and token consumption between G-Designer and other high-performing baselines on the GSM8K dataset.}\label{tab:resource}
%     \resizebox{\linewidth}{!}{
%     \begin{tabular}{c|c|c|c}
%     \toprule
%         Method & Accuracy (\%) & Token Count & \makecell{Wall-clock\\ Time} \\
%         \midrule
%         Complete Graph & 86.49 & $9.8 \times 10^6$ & 2.4h \\
%         DyLAN & 88.16 & $2.2 \times 10^7$ & 7.4h \\
%         GPTSwarm & 89.74 & $1.4 \times 10^7$ & 4.5h \\
%         G-Designer & \textbf{95.07} & $\mathbf{5.5 \times 10^6}$ & \textbf{2.6h} \\
%         \bottomrule
%     \end{tabular}}
%     \vspace{-0.6em}
% \end{table}  

\begin{table}[h]
    \centering
    \vspace{-1em}
        \caption{Efficiency analysis. We compare the training/inference wall-clock time and token consumption between \ourmethod and other high-performing baselines on the GSM8K dataset.}
    \label{tab:resource}
        \resizebox{\linewidth}{!}{
    \begin{tabular}{l|ccccccc}
        \toprule
        Method & Perf. & \makecell{\#Training\\ Token} & \makecell{\#Inference\\ Token} & \makecell{\#Overall\\ Token} & \makecell{Training\\ Time}& \makecell{Inference\\ Time}  \\
        \midrule
        Complete & 86.4 & - & $9.8 \times 10^6$ & $9.8 \times 10^6$  & - & 2.4h \\
        DyLAN & 88.1 & $9.6 \times 10^6$ & $1.3 \times 10^7$ & $2.2 \times 10^7$  & 2.8h & 4.6h  \\
        GPTSwarm & 89.7 & $5.5 \times 10^6$ & $8.4 \times 10^6$ & $1.4 \times 10^7$& 2.1h & 2.8h \\
        \midrule
        \textbf{\ourmethod} & \textbf{95.0} & $\mathbf{2.7 \times 10^5}$ & $\mathbf{8.2 \times 10^6}$ & $\mathbf{8.5 \times 10^6}$ & \textbf{0.3h} & \textbf{2.3h}  \\
        \bottomrule
    \end{tabular}}
\end{table}

\begin{figure}[!t]
  \centering
  \includegraphics[width=1\linewidth]{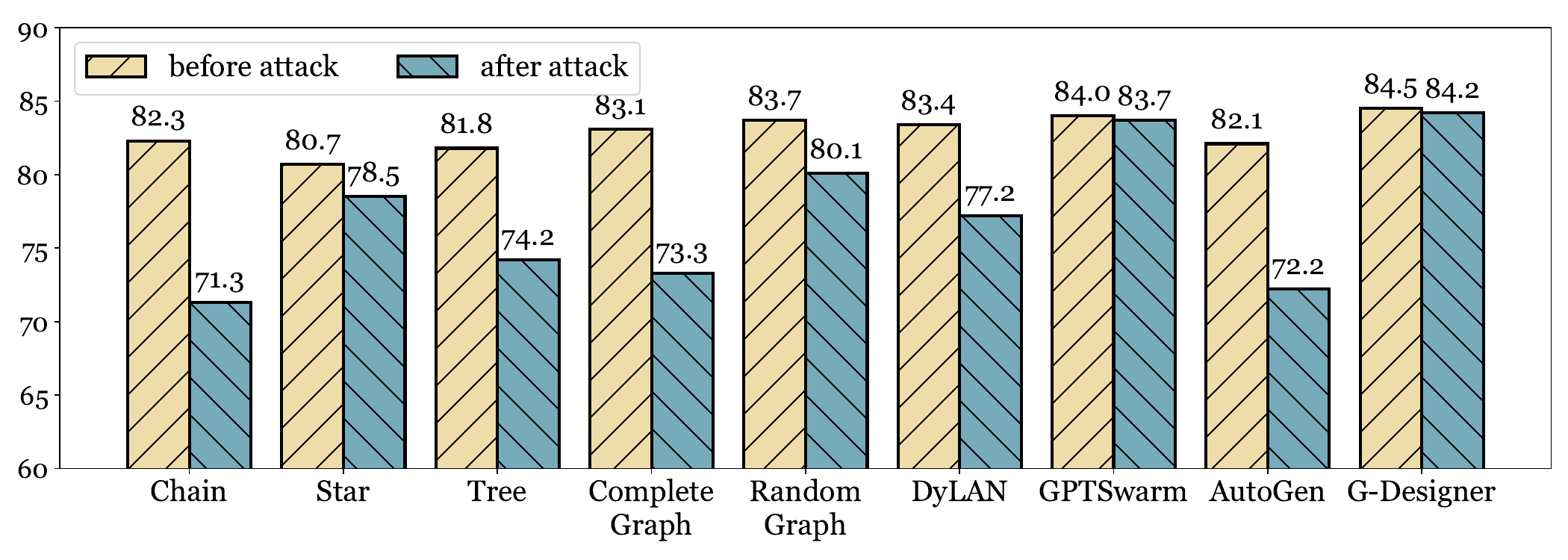}
  \vspace{-2.5em}
  \caption{We compare the accuracy (\%) of various multi-agent
frameworks before and after prompt attacks on MMLU. }
   \label{fig:robust}
   \vspace{-1.em}
\end{figure}

\vspace{-1em}
\subsection{Robustness Analysis}
\vspace{-0.7em}
% We compare the robustness of different topology designs when subjected to adversarial attacks. 
Following \citep{zhuge2024gptswarm}, we simulate a system prompt attack on one of the five agents. As seen in \Cref{fig:robust}, many trivial structures, such as chain or complete graph, experience significant performance degradation under partial system attacks, with drops as high as $11.0\%$. Among more sophisticated structures, GPTSwarm, benefiting from its specialized node optimization mechanism, only suffers a minor $0.3\%$ accuracy decline. However, other methods fare less well, with DyLAN and AutoGen showing accuracy drops of $6.2\%$ and $9.9\%$, respectively. Remarkably, \ourmethod demonstrates exceptional robustness against adversarial attacks, maintaining nearly identical performance pre- and post-attack. This resilience can be attributed to its agent encoding capability, which, during optimization, can detect malicious inputs and prune the corresponding edges.

\vspace{-0.5em}
\subsection{Framework Analysis}\label{sec:analysis}

% \begin{figure}[!t]
%   \centering
%   \includegraphics[width=1\linewidth]{img/sensi.pdf}
%   \vspace{-2.em}
%   \caption{(\textit{Left})) Sensitivity analysis on the number of agents $N$. (\textit{Right}) Ablation study of two regularizations under clean and adversarial attack settings, tested on MMLU benchmark.}
%    \label{fig:sensi}
%    \vspace{-1.1em}
% \end{figure}
% \vspace{-0.1em}

\begin{table}[!t]
    \centering
        \caption{Ablation study  tested on MMLU benchmark.}
    \label{tab:variant_performance}
   \resizebox{\linewidth}{!}{
    \begin{tabular}{l|cc|cc}
        \toprule
        \multirow{2}{*}{Variant} & \multicolumn{2}{c|}{MMLU} & \multicolumn{2}{c}{GSM8K} \\
        \cline{2-5}
        & Clean & Attack & Clean & Attack \\
        \midrule
        vanilla \ourmethod & 84.5 & 84.2 & 95.0 & 92.5 \\
        \midrule
        \textit{w/o} SR & 84.1 & 83.2 & 94.4 & 90.7 \\
        \textit{w/o} Anchor & 84.0 & 83.8 & 94.7 & 92.0 \\
        \textit{w/o} $\operatorname{NodeEncoder(\cdot)}$ & 83.2 & 82.4 & 92.8 & 87.4 \\
        \textit{w/o} $v_\text{task}$ & 81.3 & 82.0 & 90.3 & 87.7 \\
        \bottomrule
    \end{tabular}
    }
\vspace{-1.5em}
\end{table}

% \paragraph{\textbf{Sensitivity Analysis}} We compare the performance of the chain structure, GPTSwarm, and \ourmethod across varying numbers of agents $N$. As shown in \Cref{fig:sensi} (\textit{Left}), with the increase in agent count, the simple chain-style structure exhibits marginal performance improvements and poor scaling capacity. In contrast, \ourmethod demonstrates a stronger emergent capability, where the involvement of more agents leads to notable performance gains.
\vspace{-0.5em}

\paragraph{{Ablation Study.}} We report results for two variants of ourmethod: \textbf{(1) \textit{w/o} SR}, which removes the sparsity regularization in \Cref{eq:low_rank}, \textbf{(2) \textit{w/o} Anchor}, which excludes the anchor structure $\mathbf{A}_{anchor}$, \textbf{(3) \textit{w/o} NodeEncoder}, removing node encoder in \Cref{eq:node-encoder}, and \textbf{(4) \textit{w/o} $v_\text{task}$} in \Cref{eq:task_graph}.  As shown in \Cref{tab:variant_performance}, removing the task virtual node disrupts \ourmethod's task-adaptiveness, leading to the most significant performance drop. 
The removal of $\mathbf{A}_{anchor}$ consistently leads to performance degradation, while the absence of sparsity regularization makes the system more vulnerable to adversarial attacks.

\vspace{-1.4em}
\paragraph{Discussion on anchor topology.} 
Given that \ourmethod is initialized with the anchor topology $\mathbf{A}_{\text{anchor}}$ introduced in \Cref{sec:multi-agent-network}, one may question whether the performance gains of \ourmethod primarily stem from $\mathbf{A}_{\text{anchor}}$ itself.  
In response, we emphasize that the anchor topology corresponds to the simple Chain structure in \Cref{tab:rq1_performance}, where \ourmethod achieves substantial improvements over it, specifically $9.50\%\uparrow$ on GSM8K and $8.44\%\uparrow$ on SVAMP.  
Thus, we assert that the superior performance of \ourmethod is predominantly attributed to its adaptive topology design rather than the anchor topology itself.

\vspace{-0.5em}
\section{Conclusion}\label{sec:conclusion}
\vspace{-0.5em}
In this paper, we present the LLM-based Multi-agent Communication Protocol (MACP), which aims to provide insightful guidance for designing complex multi-agent systems. Furthermore, we propose an effective, adaptive, and robust LLM-powered multi-agent communication graph designer, termed \ourmethod, to facilitate the automated design of collaborative AI systems. \ourmethod is highly task-aware, dynamically crafting compact and robust communication topologies based on task complexity. We hope that \ourmethod will inspire future research on self-organizing and self-evolving collective intelligence.

% Acknowledgements should only appear in the accepted version.
% \section*{Acknowledgements}

% \textbf{Do not} include acknowledgements in the initial version of
% the paper submitted for blind review.

% If a paper is accepted, the final camera-ready version can (and
% usually should) include acknowledgements.  Such acknowledgements
% should be placed at the end of the section, in an unnumbered section
% that does not count towards the paper page limit. Typically, this will 
% include thanks to reviewers who gave useful comments, to colleagues 
% who contributed to the ideas, and to funding agencies and corporate 
% sponsors that provided financial support.

\section*{Impact Statement}

\paragraph{Ethical impacts.} We confidently affirm that our paper is free of ethical concerns across its motivation, design, experiments, and data usage. The proposed \ourmethod method aims to advance the fields of multi-agent systems and communication topologies design automation, contributing positively and responsibly to the scientific community.
\vspace{-1.5em}
\paragraph{Expected societal implications.} Our paper presents a significant advancement in multi-agent systems. \ourmethod addresses the challenge of choosing the right communication topology for specific tasks. \ourmethod not only provides a practical solution for multi-agent deployment but also paves the way for future studies on collective intelligence systems. Furthermore, as long as the base LLM used is aligned with human values, our system will not generate harmful content. 

% In the unusual situation where you want a paper to appear in the
% references without citing it in the main text, use \nocite
\nocite{langley00}

\bibliography{example_paper}
\bibliographystyle{icml2025}

%%%%%%%%%%%%%%%%%%%%%%%%%%%%%%%%%%%%%%%%%%%%%%%%%%%%%%%%%%%%%%%%%%%%%%%%%%%%%%%
%%%%%%%%%%%%%%%%%%%%%%%%%%%%%%%%%%%%%%%%%%%%%%%%%%%%%%%%%%%%%%%%%%%%%%%%%%%%%%%
% APPENDIX
%%%%%%%%%%%%%%%%%%%%%%%%%%%%%%%%%%%%%%%%%%%%%%%%%%%%%%%%%%%%%%%%%%%%%%%%%%%%%%%
%%%%%%%%%%%%%%%%%%%%%%%%%%%%%%%%%%%%%%%%%%%%%%%%%%%%%%%%%%%%%%%%%%%%%%%%%%%%%%%
\newpage
\appendix
\onecolumn
\section{Algorithm Workflow}

\begin{algorithm}[!ht]
\caption{Designing workflow of \ourmethod}\label{alg:algo}
\Input{Input query $\mathcal{Q}$, Graph auto-encoder $f_v$ composed of encoder $q(\cdot)$ and decoder $p(\cdot)$ (parameterized by ${\Theta}_e$ and $\Theta$), learning rate $\alpha$ }

\For{$\text{query}$ d in $\{1,2,\cdots,D'\}$}{

\textcolor{blue}{\tcc{Establish multi-agent network}}

\For{$\text{node}$ i in $\{1,2,\cdots,N\}$}{
$\mathbf{x}_i \leftarrow \operatorname{NodeEncoder}\left(\mathcal{T}(\texttt{Base}_i), \texttt{Role}_i, \mathcal{T}(\texttt{Plugin}_i)\right)$
}
Obtain agent embeddings $\mathbf{X}_{agent}\leftarrow [\mathbf{x}_1, \mathbf{x}_2,\cdots,\mathbf{x}_N]^\top$

Obtain task-specific node $x_{task} \leftarrow \operatorname{NodeEncoder}(\mathcal{Q}_d)$

Set an anchor topology $\mathbf{A}_{anchor}$  \textcolor{blue}{\tcp{ In our experiments, the anchor topology is simply set as the chain structure}}

Obtain a task-specific multi-agent network $\Tilde{\mathcal{G}}= \bigl(\begin{bmatrix}
   \mathbf{X}_{agent} \\
   \mathbf{x}_{task}^\top
   \end{bmatrix}, \mathbf{A}_{anchor}\bigl)$ \tcp{Note that $\mathbf{A}_{anchor}$ here contains bidirectional edges added by the task node $v_{task}$}

\textcolor{blue}{\tcc{Design communication topology}}

Encode $\Tilde{\mathcal{G}}$ into latent agent representations $\mathbf{H}_{agent}$:  $q(\mathbf{H}_{agent}\;|\;\Tilde{\mathbf{X}},\mathbf{A}_{anchor})=\prod_{i=1}^{N}q(\mathbf{h}_i\;|\;\Tilde{\mathbf{X}},\mathbf{A}_{anchor})$

Decode (phase 1) and obtain the sketched graph $\mathbf{S}$: $p_s(\mathbf{S} \;|\; \mathbf{H}_{agent}) = \prod_{i=1}^N\prod_{j=1}^N p_s(\mathbf{S}_{ij}\;|\;\mathbf{h}_i,\mathbf{h}_j,\mathbf{h}_{task};\Theta_d),$

Decode (phase 2) and obtain the communication graph $\mathcal{G}_{com} = (\mathcal{V}, \mathcal{E}_{com}), \mathcal{E}_{com} = \{(i,j)\;|\;\Tilde{\mathbf{S}}_{ij}\neq 0 \wedge (i,j)\in\mathcal{E}\})$

\tcc{Guide multi-agent system collaboration}

\For{$\text{iteration}$ t in $\{1,2,\cdots,K\}$}{
\For{$\text{node}$ i in $\phi(\mathcal{G}_{com})$}{
Agent $v_i$ generates $\mathcal{R}_i^{(t)} \leftarrow v_i(\mathcal{P}^{(t)}_\text{sys}, \mathcal{P}^{(t)}_\text{usr}),\; \mathcal{P}^{(t)}_\text{usr}=\{\mathcal{Q}, \cup_{v_j \in \mathcal{N}_\text{in}(v_i)} \mathcal{R}_j^{(t)}\}$
}
\tcc{Aggregate solution}
$a^{(t)} \leftarrow \operatorname{Aggregate}(\mathcal{R}_1^{(t)}, \mathcal{R}_2^{(t)},\cdots, \mathcal{R}_N^{(t)})$
}
\tcc{Update \ourmethod parameters}
$\Theta^{d+1} \leftarrow \Theta^d - \alpha \nabla_{\Theta^d}\mathcal{L}_{\ourmethod}$

}
\end{algorithm}

\section{Dataset Statistics}

We conclude the dataset statistics in \Cref{tab:dataset_stats}.

\begin{table}[h]
    \centering
    \caption{{Dataset descriptions and statistics.}}
    \label{tab:dataset_stats}
    \begin{tabular}{llcccl}
        \toprule
        \textbf{Category} & \textbf{Dataset} & \textbf{Answer Type} & \textbf{Metric} & \textbf{\#Test} & \textbf{License} \\
        \midrule
        \multirow{1}{*}{General reasoning} & MMLU & Multi-choice & Acc. & 153 & MIT License \\
        \midrule
        \multirow{4}{*}{Math reasoning} 
        & GSM8K & Number & Acc. & 1,319 & MIT License \\
        & MultiArith & Number & Acc. & 600 & Unspecified \\
        & SVAMP & Number & Acc. & 1,000 & MIT License \\
        & AQuA & Multi-choice & Acc. & 254 & Apache-2.0 \\
        \midrule
        \multirow{1}{*}{Code generation} & HumanEval & Code & Pass@1 & 164 & MIT License \\
        \bottomrule
    \end{tabular}
\end{table}

\section{Case Study}

\begin{figure*}[!t]
  \centering
  \includegraphics[width=1\linewidth]{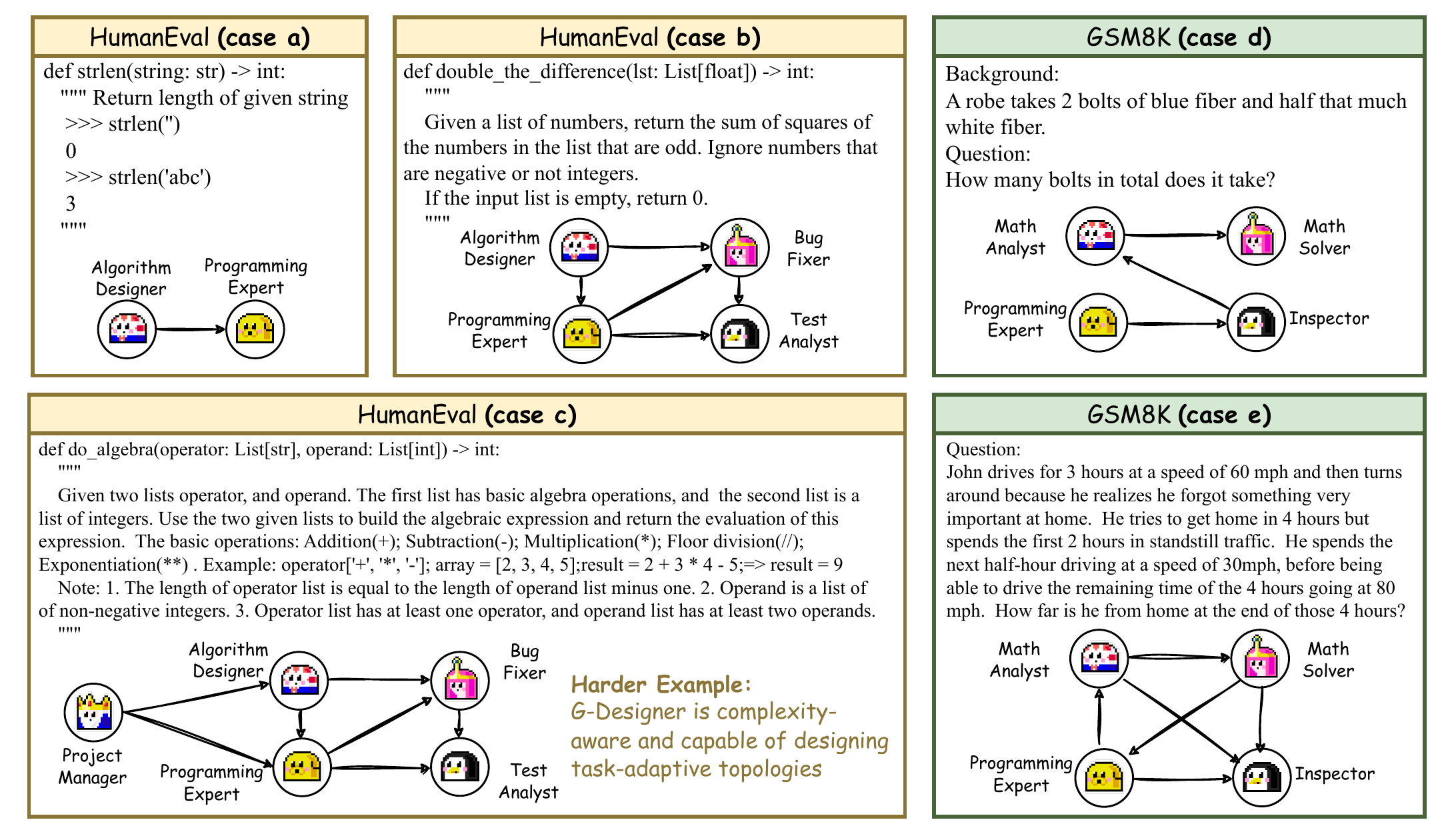}
  \vspace{-2.4em}
  \caption{Case study of the communication topologies designed by \ourmethod on HumanEval and GSM8K benchmarks. }
   \label{fig:case_study}
   \vspace{-1.3em}
\end{figure*}

\Cref{fig:case_study} visualizes the different topologies designed by \ourmethod for varying query difficulties on the HumanEval and GSM8K benchmarks.

\section{Supplementary Results}

\begin{table}[h]
    \centering
    \caption{The GPU cost of \ourmethod with increasing number of agents.}\label{tab:gpu}
    \begin{tabular}{l|cccc}
    \toprule
        \#Agents & 5 & 50 & 100 & 1000 \\
        \midrule
        Memory (GB) & 2.7 & 2.9 & 3.0 & 3.8 \\
        \bottomrule
    \end{tabular}
\end{table}  

% \begin{table}[h]
%     \centering
%         \caption{Comparison of accuracy, time, token consumption, and cost across different agent configurations.}
%     \label{tab:agent_scale}
%     \resizebox{\linewidth}{!}{
%     \begin{tabular}{c|cc|cc|cc}
%         \toprule
%         Metric & \multicolumn{2}{c|}{5 Agents} & \multicolumn{2}{c|}{10 Agents} & \multicolumn{2}{c}{20 Agents} \\
%         \cline{2-7}
%         & Complete Graph & \ourmethod & Complete Graph & \ourmethod & Complete Graph & \ourmethod \\
%         \midrule
%         Accuracy (\%) & 71.90 & \textbf{73.20} & 72.16 & \textbf{74.51} & 72.51 & \textbf{77.82} \\
%         Time (min) & 16.9 & \textbf{19.3} & 34.2 & \textbf{36.0} & 66.5 & \textbf{68.9} \\
%         \#Prompt Tokens & 545,984 & \textbf{452,329} & 1,669,451 & \textbf{885,332} & 5,648,834 & \textbf{1,852,538} \\
%         Cost (USD) & 0.7 & \textbf{0.6} & 2.2 & \textbf{1.3} & 7.4 & \textbf{2.7} \\
%         \bottomrule
%     \end{tabular}}
% \end{table}

\begin{table}[h]
    \centering
        \caption{Comparison of accuracy, time, token consumption, and cost across different agent configurations. We use the MMLU benchmark and \llmname{gpt-3.5-turbo} as the base LLM.}
    \label{tab:agent_scale}
    \begin{tabular}{c|cccc}
        \toprule
        \#Agents & 5 & 10 & 20 \\
        \midrule
        \textbf{Chain} & & & \\
        Accuracy (\%) & 70.59 & 71.24 & 71.98 \\
        Time (min) & 15.73 & 30.20 & 56.18 \\
        \#Prompt Tokens & 351,802 & 702,164 & 1,378,328 \\
        Cost (USD) & 0.5228 & 1.0434 & 2.0482 \\
        \midrule
        \textbf{Complete Graph} & & & \\
        Accuracy (\%) & 71.90 & 72.16 & 72.51 \\
        Time (min) & 16.85 & 34.21 & 66.47 \\
        \#Prompt Tokens & 545,984 & 1,669,451 & 5,648,834 \\
        Cost (USD) & 0.7161 & 2.1770 & 7.3662 \\
        \midrule
        \textbf{GPTSwarm} & & & \\
        Accuracy (\%) & 72.55 & 73.86 & 75.38 \\
        Time (min) & 62.14 & 186.86 & 412.18 \\
        \#Prompt Tokens & 3,055,236 & 9,048,465 & 30,317,341 \\
        Cost (USD) & 4.2190 & 12.4961 & 41.4235 \\
        \midrule
        \textbf{\ourmethod} & & & \\
        Accuracy (\%) & \textbf{73.20} & \textbf{74.51} & \textbf{77.82} \\
        Time (min) & 19.26 & 36.04 & 68.89 \\
        \#Prompt Tokens & 452,329 & 885,332 & 1,852,538 \\
        Cost (USD) & 0.6036 & 1.2768 & 2.6713 \\
        \bottomrule
    \end{tabular}
\end{table}

%%%%%%%%%%%%%%%%%%%%%%%%%%%%%%%%%%%%%%%%%%%%%%%%%%%%%%%%%%%%%%%%%%%%%%%%%%%%%%%
%%%%%%%%%%%%%%%%%%%%%%%%%%%%%%%%%%%%%%%%%%%%%%%%%%%%%%%%%%%%%%%%%%%%%%%%%%%%%%%

\end{document}